\documentclass[12pt]{iopart}

\usepackage{graphics,graphicx}
\usepackage{hyperref}
\begin{document}

\title{An analytical approach to symmetry breaking in multipole RF-traps}

\author{M. Marchenay and J. Pedregosa-Gutierrez and M. Knoop and M. Houssin and C. Champenois}

\address{Aix Marseille Univ, CNRS, PIIM,  Marseille, France}
\ead{caroline.champenois@univ-amu.fr}

\vspace{10pt}
\begin{indented}
\item[] 10 december 2020
\end{indented}

\begin{abstract}
Radio-frequency linear multipole traps have been shown to be very sensitive to mis-positioning of their electrodes, which results in a symmetry breaking and leads to extra local minima in the trapping potential as shown in \cite{pedregosa17} disturbing the operation of the trap. In this work, we analytically describe the RF-potential of a realistic octupole trap by including lower order terms to the well-established equation for a perfectly symmetric octupole trap. We describe the geometry by a combination of identified defects, characterised by simple analytical expressions. A complete equation is proposed for a trap with any electrode deviation relying on a combination of the simple cases where the defects are taken individually. Our approach is validated by comparison between analytical and numerical results for defect sizes up to 4\% of the trap radius. As described in \cite{pedregosa18}, an independent fine-tuning of the amplitude of the RF voltage applied on each electrode can be used to mitigate the geometrical defects of a realistic trap. In a different way than in \cite{pedregosa18}, the knowledge of an analytical equation for the potential allows to design the set of RF-voltages required for this compensation, based on the experimental measurement of the ion positions in the trap, without information concerning the exact position of each electrode, and with a small number of iterations.  The requirements, performances and limitations of this protocol are discussed via comparison of numerical simulations and analytical results.  
\end{abstract}

%
%
%
%
%

\section{Introduction}
The trapping of charged particles by radio-frequency (RF) electric fields has proven to be both a powerful and versatile tool for experimental exploration in physics and chemistry. Even if the linear quadrupole trap is mostly used for mass spectroscopy \cite{douglas05}, optical metrology \cite{brewer19}, quantum computing \cite{monz11} and high precision measurements \cite{alighanbari20, patra20} for its harmonic potential shape, linear multipole traps happen to be candidate of choice in the design of micro-wave atomic clocks for spacecraft navigation \cite{prestage07,burt08} or for cold collisionand chemical reaction studies \cite{gerlich07,wester09, gunther17,simpson20}. Both these applications take advantage of the lower RF driven motion amplitude which results in a lower Doppler effect and a lower kinetic energy in the collision frame, compared to the same sample in a quadrupole trap. When laser cooled in a multipole trap, the ions are expected to organise in a hollow core structure \cite{champenois09}, forming structures like rings \cite{champenois10} and tubes \cite{calvo09} that cannot be observed in conventional quadrupole traps. Recent reported experimental observations in a linear octupole trap reveal no such structure but three individual ion clouds. This is the consequence of built-in imperfections and mechanical electrodes misalignments that result in a symmetry breaking in the RF-potential arrangement of the trap \cite{pedregosa17}. A similar phenomenon has been observed in a 22-pole with the apparition of 10 zones where the ions cooled to 170~K accumulate \cite{otto09}. 
We aim at correcting the mechanical defects in the octupole trap by adjusting each electrode potential as demonstrated in \cite{pedregosa18} in a simulated experimental set-up. In this former work, a random geometry with a mean displacement of each electrodes of 1.3\% of the internal radius of the trap could be corrected on a ring large as a tenth of this internal radius after 576 iterations of trapping potential comparison and adjustment. The simulations assume that laser-cooled ions are trapped and observed in the three local minima of the pseudo-potential and the potential comparisons are based on their relative positions. The compensation leads to a trapping potential variation along this ring corresponding to 3.6~mK. To reduce drastically the number of required steps to compensate for geometrical defects, we assumed  an accurate understanding of the correlation between trap defects and experimental observations. To that purpose, we have worked on establishing an analytic equation for the instantaneous RF potential of an octupole trap with small defects. The resulting equation is entirely defined by the real position of the electrodes in the radial plan, and the electrode size. The validation of this approach is based on the comparison between the analytical potential and the one computed numerically. This comparison focuses on information that can be accessed by experiments, like the one that demonstrated the three local minima in the trapping potential \cite{pedregosa17}.  

The method used as an analytical description for the rf potential is introduced in the next section. Section \ref{s_pert_list} details how to convert the electrode position into an equation, using a basis of five identified deformations that can be superposed to describe any configuration.  Section~\ref{s_cor} explains the correction protocol that we developed, based on the previous analytical equations and a set of compensation voltages. The strategy first tested for defect compensation can then be used to create on-demand configurations of three parallel trapping zones.

\section{A 2D-octupole potential as a combination of main and perturbation terms}\label{s_pert}
The objective of this first part is to build an analytical equation for the time dependent electric potential in the radial plane of an asymmetric linear octupole trap by adding perturbations to the potential equation of an ideal octupole trap. Any lack of parallelism between the electrodes is not taken into account and we only consider a mismatch of electrode positions in a transverse plane. Using a perturbative approach is motivated by the need to have a  description of the trapping pseudo-potential whereas the defects are accurately described in the time dependent electric field. The starting point is the equation for time-dependent potentials in a perfectly symmetric multipole trap $\psi_n(x,y,t)$, written as in Eq.~\ref{eq_perfect_pot} where $V_{RF}$ is the amplitude of the RF potential applied to the electrodes, $\Omega_{RF}/2\pi$ is the radio-frequency and $K_n(x,y)$ is a 2D surface defined by the order $n$ of the multipole, $2n$ being the number of electrodes in the trap \cite{friedman82} :
\begin{equation}\label{eq_perfect_pot}
\psi_n(x,y,t)=V_{RF} K_n(x,y)\cos(\Omega_{RF}t)
\end{equation}
The $K_n(x,y)$ function is the real part (noted $U_n(x,y)$) or imaginary part (noted $V_n(x,y)$) of $(x+iy)^n/r_0^n$ where $r_0$ is the inner radius of the trap.  For an octupole ($n=4$), with  the orthogonal $(x,y)$ frame crossing  the center of 4 electrodes out of 8, $K_n(x,y)$ is $U_4(x,y)$ :
\begin{equation}\label{eq_U4}
U_4(x,y)=\left(x^4-6x^2y^2+y^4\right)/r_0^4.
\end{equation}
This equation assumes that the electrode shape fits the iso-potential lines, and imposes a hyperbolic section for the electrodes. However for practical reasons, the use of circular electrodes is favoured with little modification of the potential in the center of the trap if the ratio  between the electrode radius $r_d$ and the inner radius of the trap $r_0$ matches a specific value that depends on the multipole trap order (0.333 for octupole traps and  1.1451 for  quadrupole traps)\cite{reuben96}. Cylindrical electrodes only are within the scope of this article. 

With the analytical or numerical description of the RF-potential $\psi_n(x,y,t)$, we compute the pseudo-potential using the method first demonstrated for a quadrupole trap   in  \cite{dehmelt67} and extrapolated to a multipole trap \cite{gerlich92, champenois09}. In this approach, the  particles are trapped in a static potential which corresponds to the time-average of the squared local RF electric field. One can show that the static pseudo-potential can be computed by
\begin{equation}\label{eq_pseusoV}
V^*(\mathbf{r})=\frac{q^2 \mathbf{E_0}^2(x,y)}{4 m \Omega_{RF}^2}.  
\end{equation}
with $\mathbf{E_0}(x,y)$ the local amplitude of the RF-electric field, defined as $-\mathbf{\nabla}(V_{RF} K_n(x,y))$. The experimental signature of a symmetry breaking in a $2n$-pole trap is the accumulation of trapped ions into $(n-1)$ local minima, as was observed for an octupole trap \cite{pedregosa17} and a 22-pole trap \cite{otto09}. This signature in the pseudo-potential approximation means that the RF-electric field $E_0(x,y)$ has $(n-1)$ non-degenerated $(x,y)$ zeroes. This can happen only if the $n$-order $(x,y)$ polynomial  RF-potential $\psi_n$ includes terms of lower orders than $n$. These lower order terms are cancelled when the $n$ mirror symmetries are obeyed. 

Based on the previous analysis, we build the expression of the perturbative term as a sum of lower order terms. Insight on this decomposition method can be gained by starting with the simpler case of a symmetric quadrupole potential with an inner radius~$r_0$. We call $U_2$ the $K_2$ polynomial if the axis frame is chosen crossing the center of the electrodes, and $V_2$ if it is equidistant to their center. As shown by Eq.~\ref{eq_U2} and \ref{eq_V2} they can be expressed as product and combination of the two dipole potential basis terms $U_1(x,y)=x/r_0$ and $V_1(x,y)=y/r_0$. 
\begin{eqnarray}
U_2(x,y)&=&\left(x^2-y^2\right)/r^2_0=U_1^2(x,y)-V_1^2(x,y)   \label{eq_U2}\\
V_2(x,y)&=&\left(2xy\right)/r^2_0=2U_1(x,y)V_1(x,y)   \label{eq_V2}
\end{eqnarray}
In the same manner, and within the same convention as for $(U_2, V_2)$ for the relation of the frame orientation,  the equation for the octupolar potential can be seen as a combination of quadrupole terms and thus, dipole terms (Eq.~\ref{eq_U4}).
\begin{eqnarray}
U_4(x,y)&=&U_2^2(x,y)-V_2^2(x,y)   \label{eq_U4}\\
V_4(x,y)&=&2U_2(x,y)V_2(x,y)   \label{eq_V4}
\end{eqnarray}

We assume that the lower order terms responsible for extra zeroes  in the octupole potential equation  can be mapped onto quadrupole and dipole contributions, as it is the case for the symmetric geometry. We therefore propose $U_R(x,y)$ as an equation for an asymmetric octupole potential function $K_4(x,y)$, with a frame axis set crossing the electrode center when it is "perfect" 
 \begin{equation}\label{eq_U}
 U_R(x,y)=h_0 [U_4(x,y)-W(x,y)]
\end{equation}
 with the perturbation term
\begin{equation}\label{eq_W}
W(x,y)=a_1U_2(x,y)+a_2V_2(x,y)+a_3U_1(x,y)+a_4V_1(x,y).
\end{equation}
The global $h_0$ scaling parameter accounts for the impact of the radius ratio $r_d/r_0$ (see \ref{s_rayon} for $h_0$ values). Symmetry breaking could also be responsible for higher order terms that modify the profile of the potential for large enough distance to the trap center. We neglect these contributions for this work.

\begin{figure}[htbp]
\begin{center}
\includegraphics[width=10cm]{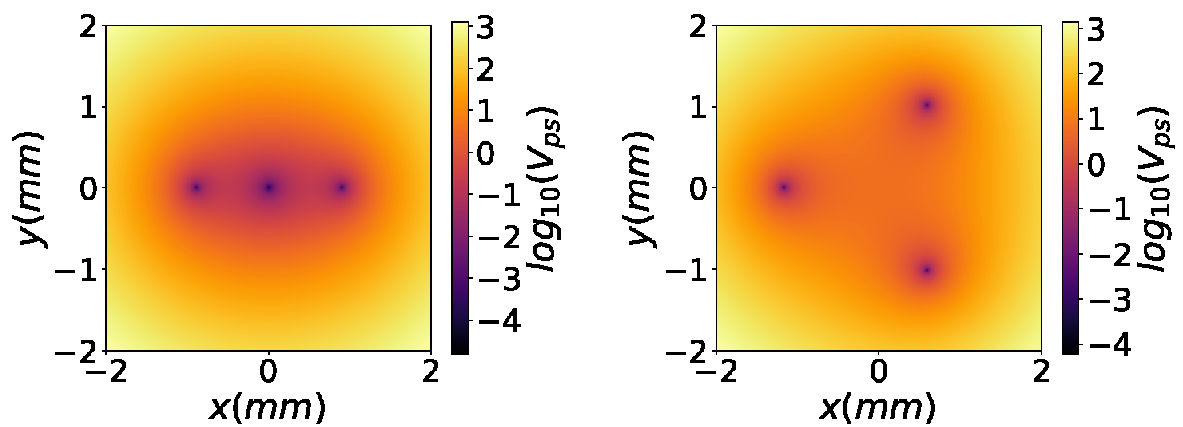}
\caption{ Pseudo-potential shape for two simple cases of added perturbation (a) Quadrupole term with  $a_1=0.1, \ a_2=a_3=a_4=0$, and  (b) Dipole term with $a_3=-0.1, \ a_1=a_2=a_4=0$, for a $r_0=4$~mm trap \label{fig_demo}}
\end{center}
\end{figure}

A single perturbation has a clear signature on the pseudo-potential, like shown on Fig~\ref{fig_demo}. In the first example, a purely quadrupole perturbation, coded by ($a_1=0.1, \ a_2=a_3=a_4=0$)  creates three equally spaced potential minima aligned along the $x$-axis (for  $a_1<0$ the minima are aligned along the $y$-axis). Using the equations, one can show that the distance between the minima is $r_0\sqrt{|a_1/2|}$ and this perturbation was already proposed in \cite{marciante11} to trap three parallel ion strings and control their separation. The second example adds a dipole term perturbation coded by ($a_3=-0.1, \ a_1=a_2=a_4=0)$, which creates  a near-equilateral triangle organisation of the minima in the potential. Even if the zeroes are not easily accessible to an analytic solution, one can show that  one of the minima sits on the $x$-axis and the two other ones are equidistant from this $x$-axis. These individual cases allow us to associate a linear configuration of potential minima to a quadrupole perturbation and a triangle configuration to a dipole perturbation.

\section{From electrode position to perturbation terms for the potential equation}\label{s_pert_list}
The next step of our demonstration is to quantify the weights $a_i$ on each quadrupole and dipole term in accordance with the positions of the  trap electrodes. We aim to define a parameter set independent of an arbitrary definition frame to establish an unambiguous characterisation of the trap deformations. In order to build a self-consistent description of the trap geometry, the electrode configuration is described by the composition of five deformation classes, based on   the superposition of two  sets of electrodes. We  call $S$ the 4-electrode set aligned with the $(x;y)$  frame  and $T$ the complementary set turned by  a $\pi/4$ angle from this frame. The five deformations that we review in the following are {\it Compression}, {\it Sliding} and {\it Splitting} which keep the orthogonality of the sets and {\it Rotation} and {\it Shearing} which are angular deformations.  The formers are  straightforward to turn into equations, but angular deformations make the problem more complex by coupling the axis used for the analytical expression. 

For each deformation class, we review   its physical origin, its relevant parameters and give as a first approach a simplified equation of the potential when the trap is deformed by this defect alone. The impact of the radius ratio $r_d/r_0$ is simplified through this section by the use of coefficients $h_{0,p,l,c,h}$  assumed constant for a given radius ratio. The calculation of these coefficients for different radius ratio is addressed in  \ref{s_rayon}. The full equation accounting for several coupled deformations is detailed in the next part.  The validation of the proposed defect description is based on experimentally accessible data. In an experiment, the trapping potential can be observed by the fluorescence emitted by the ions located in the potential minima.  By analogy, with our numerical experiment,   we compare the pseudo-potential created by the analytical RF potential given by Eq.~(\ref{eq_U},\ref{eq_W})   to the one computed by CPO \cite{CPO}, a code that solves the Laplace equation based on a boundary condition imposed by the surface of the electrodes. The dimensions of the trial trap are $r_0=4$~mm and $r_d=1.5$~mm and this geometry is discretised with pixel size equal to $r_0/1000$ (4~$\mu$m). To be relevant, this comparison must not rely on an artificial frame center. To avoid this issue, each arrangement of the potential minima is referenced to the center of the triangle or line they form.

\subsection{Compression}
\begin{figure}[htbp]
\begin{center}
\includegraphics[width=8cm]{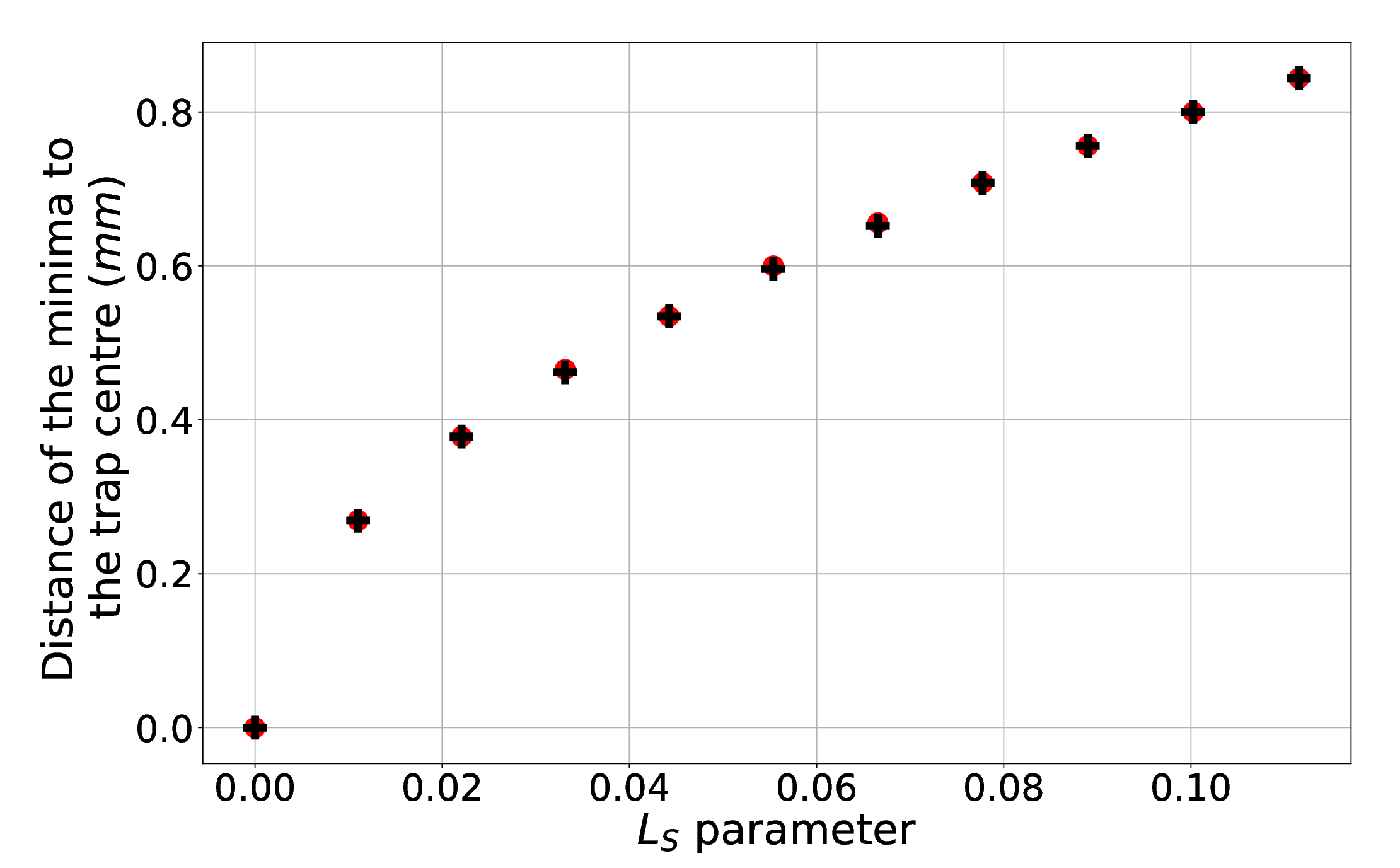}
\caption{Distance to the trap centre of the two outside minima vs the compression reduced parameter $L_{S}$  increased from 0 to 0.11 by step of 0.011. The black points are the results computed with the CPO software and the red points are calculated from the analytical equation Eq.~(\ref{eq_C}).    The size of the error bar corresponds to the pixel size for the representation, which is 3.64~$\mu$m. Trap inner radius $r_0=4$~mm and electrode radius $r_d=1.5$~mm}\label{fig_C}
\end{center}
\end{figure}

A compression describes a situation where there is a modification, in at least one of the quadrupole sets, of the distance between two electrodes facing each other, without displacement of the center of each line joining them (see Fig.~\ref{fig:motif} for a pattern). The CPO calculations show that the minima formed by such a defect arrange on a line and remain centred on the center of the trap, exactly like when a quadrupole contribution is added to the instantaneous potential (Eq.~\ref{eq_U2} and Fig.~\ref{fig_demo}). To write the perturbation $W_{c}$ induced on the instantaneous potential by such a defect, we define $L_S$ (resp $L_T$), the length difference between the two directions within the $S$-quadrupole set (resp $T$-set), normalised by the averaged inner radius $\overline{r}_0$, such that
\begin{equation}\label{eq_C}
W_{c}=h_c\left[L_SU_2(x,y)+L_TV_2(x,y)\right].
\end{equation}
Here, $h_c$ is adjusted to match the minima positions between the analytic expression and the CPO calculations and which only depends on the radius ratio $r_d/r_0$, like discussed in \ref{s_rayon}. For our trap dimensions $h_c=0.820$ and the perturbation code is ($a_1=h_cL_S$, $a_2=h_cL_T$, $a_3=a_4=0$). The sign assigned to the length modifications must be consistent with the convention chosen for the $U_2$ and $V_2$ contributions. To that end, $L_S$ is positive if the length along the $x$-axis is larger than the one along the $y$-axis and $L_T$ is positive if the length along the $(x-y)$-axis is larger than the one along the $(x+y)$-axis. The comparison between analytical and numerical calculations of the pseudo-potential minima, shown on Fig.~\ref{fig_C}, confirms that this analytical representation is relevant for a surface grid of 4~$\mu$m pixel size, for a compression reaching 8\% of the averaged  radius $(r_0+r_d)$. This maximum compression corresponds to the situation where two facing electrodes would be mis-positioned by 220~$\mu$m each. 

For the particular case of an equal compression along the two  directions defining the same quadrupole set, there is no extra minima to be observed. Indeed, in that case $L_S$ and $L_T$ remain null but the averaged inner radius  is modified, changing the strength of the trapping potential without changing the number of local minima. In the basic equations, the normalisation factor is rescaled by changing $r_0$ to $\bar{r_0}$

\subsection{Sliding}\label{sliding}
\begin{figure}[htbp]
\begin{center}
\includegraphics[width=8cm]{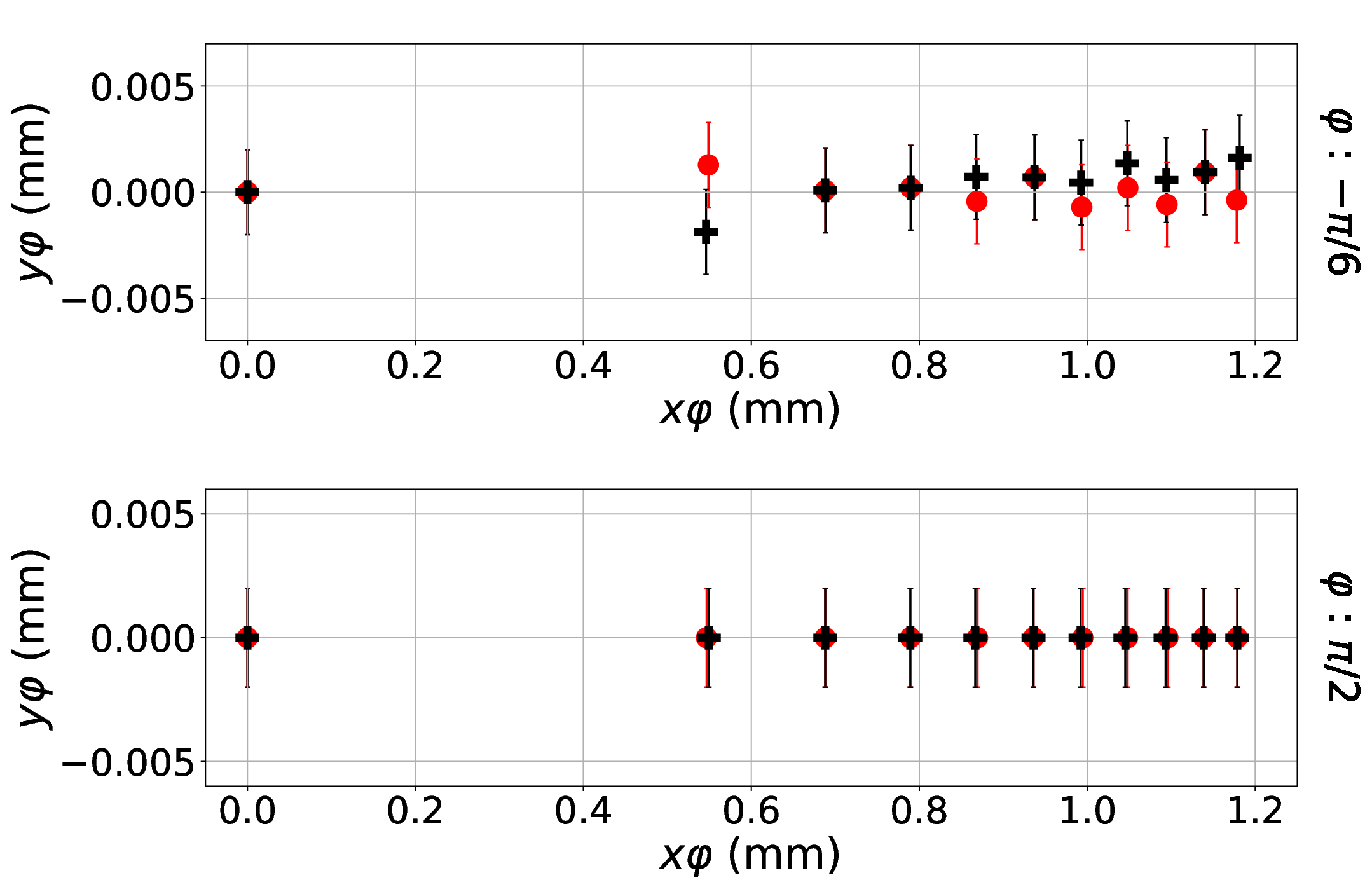}
\caption{Positions of two out of the three minima along the line over which they organise for small enough sliding of the electrodes set over the $y$-axis in the $S$ quadrupole set. The reduced sliding parameter $y^T_{l}$ is increased from 0 to $4\times 10^{-2}$ by step of $4\times 10^{-3}$. The black points are the results computed with the CPO software and the red points are calculated from the analytical equation (\ref{eq_Sd}).   The error bars correspond to the pixel size (4~$\mu$m). The maximum $y^T_{l}$ value corresponds to a common translation of two facing electrodes of 220~$\mu$m for a trap inner radius $r_0=4$~mm and electrode radius $r_d=1.5$~mm.}\label{fig_Sl}
\end{center}
\end{figure}

Like the compression, a sliding is a deformation described within one quadrupole set. There, and contrary to compression, the distance between facing electrodes is conserved but the lines joining them do not cross in their center (noted $M^S$ in the $S$-set and $M^T$ in the $T$-set, see also Fig.~\ref{fig:motif}). The characteristic parameters $(x^S_{l},y^S_{l})$ and/or $(x^T_{l},y^T_{l})$ are defined by the positions of these lines centre relatively to $M^S(X^{S},Y^{S}$) and $M^T(X^{T},Y^{T}$), the crossing points of these two lines in the $S$ and $T$-quadrupole sets. The definition of $M^S$ and $M^T$ is necessary in a general case, but for a sliding alone the two points are not separated. For a sliding in the $S$-set, $x^S_l$ (resp $y^S_l$) is the sum of the $x$ (resp $y$) coordinates relatively to $M^S$ of the two lines centres of this set. The same definition is transposed in the $T$-set. These coordinates are normalised by the distance between the centres of the facing electrodes. This kind of deformation is  responsible for a triangular organisation of the potential minima, like induced by a dipole term. With our previous convention, this perturbation is  then coded as  $a_1=a_2=0$, $a_3=h_l(x^S_{l}+x^T_{l})$ and $a_4=h_l(y^S_{l}+y^T_{l})$.
The comparison between the pseudo-potential computed by CPO  and the one deduced from the instantaneous potential equation
\begin{equation}\label{eq_Sd}
W_{l}=h_l\left[\left(x^S_{l}+x^T_{l}\right)U_1(x,y)+\left(y^S_{l}+y^T_{l}\right)V_1(x,y)\right]
\end{equation}
is shown on Fig.~\ref{fig_Sl} with the adjusted value $h_l=2.566$ and for the particular case of a sliding up to 4\% of $r_0$ in the vertical pair of electrodes of the $S$-quadrupole set.
This comparison shows that a sliding up to 4\% of $(r_0+r_d)$ can be completely taken into account by an extra dipole term.

\subsection{Splitting}
\begin{figure}[htbp]
\begin{center}
\includegraphics[width=8cm]{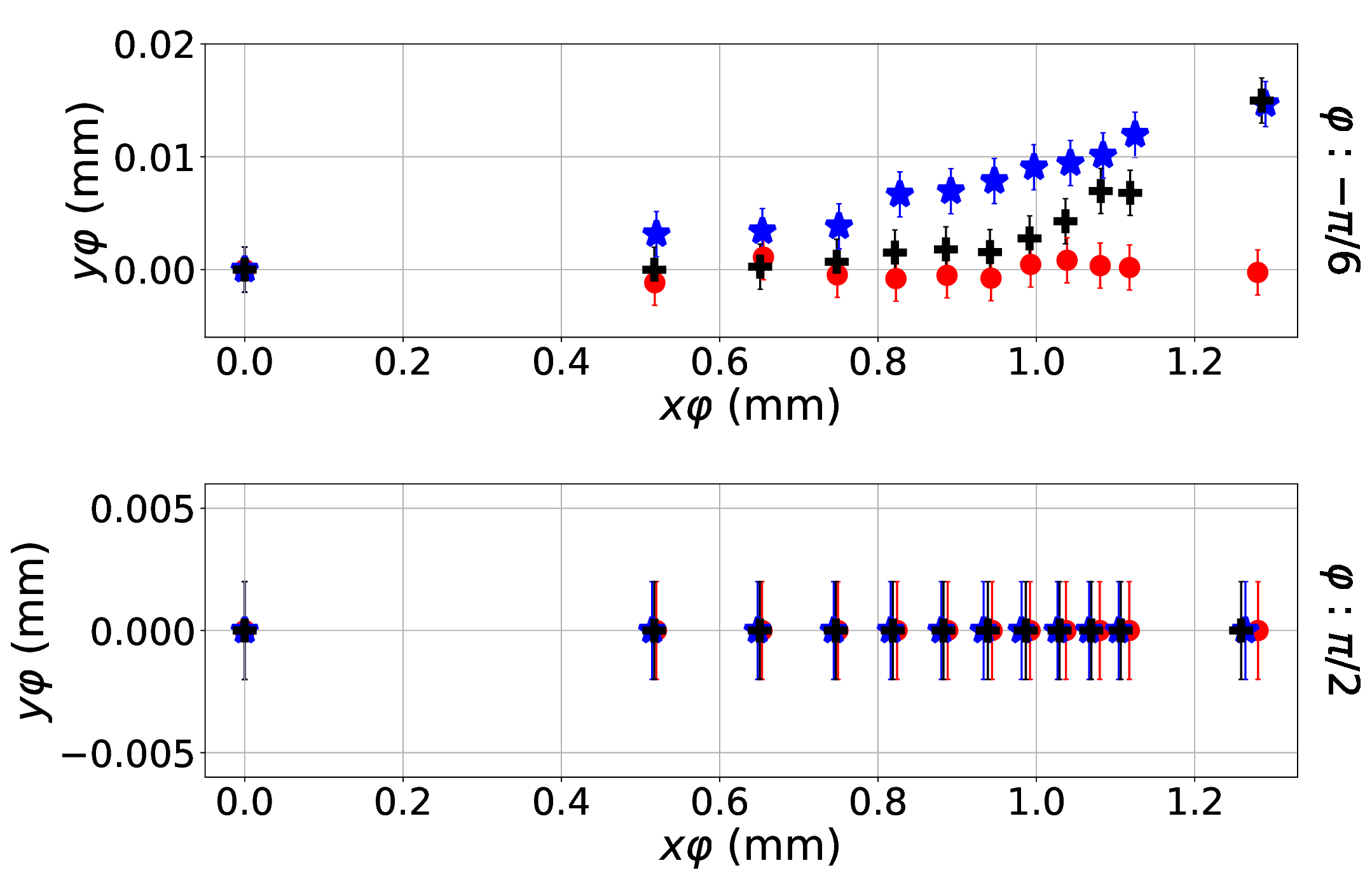}
\caption{ \label{fig_Sp} Position of two out of the three minima along the line over which they organise for small enough splitting between the $T$ and $S$ quadrupole sets. The $y$ oriented splitting is measured by the reduced  parameter $y_0$, increased from 0 to $5.5\times 10^{-2}$ by step of $5.5\times 10^{-3}$. The black crosses are the results computed with the CPO software and the red dots are calculated from the analytical Eq.~(\ref{eq_Sp}). The blue stars give the minimum position calculated with the corrected equation Eq.~(\ref{eq_Spb}). The error bars correspond to the pixel size  (4~$\mu$m). The maximum $y_0$ value corresponds to a global displacement of one quadrupole set of electrodes compared to the other one of 220~$\mu$m for a trap inner radius $r_0=4$~mm and electrode radius $r_d=1.5$~mm.}
\end{center}
\end{figure}

The Splitting defect class corresponds to the separation of the centre of the $S$ and $T$ quadrupole sets, without any deformation of each set. The relevant parameters ($x_0,y_0$) are the relative position of the $T$ set centre relatively to the $S$ one, normalised to $r_0$. The pattern formed by the three minima is similar to the one showed on Fig.~\ref{fig_demo}.b and for  small enough splittings, each minimum settles along a direction that we use as a reference to plot their positions on Fig.~\ref{fig_Sp},  for two out of three minima, because of the symmetry of the figure. This geometric configuration is similar to the one induced by a dipole perturbation and can be reproduced by adding a dipole term $W_{p}^1$ to $U_4(x,y)$ such that
  \begin{equation}\label{eq_Sp}
W_{p}^1=h_p\left[x_0 U_1(x,y)+y_0 V_1(x,y)\right].
\end{equation}
With the adjusted coefficient $h_p=1.586$ and within the convention of Eq.~\ref{eq_W}, this perturbation is coded by $a_1=a_2=0$, $a_3=h_p x_0$ and $a_4=h_p y_0$. The comparison between  the pseudo-potential minima resulting from this analytical description and from the CPO calculations  are shown on Fig.~\ref{fig_Sp} for a splitting between the $T$ and $S$ center along the $y$ axis. It shows a slight mismatch, which increases with the size of the defect. A better fit for larger defects can be reached by adding a correcting term of the quadrupole kind and the total equation for a splitting perturbation that we use in the following is 
  \begin{equation}\label{eq_Spb}
W_{p}=W_{p}^1+\frac{h_p^{'}}{\sqrt{x_0^2+y_0^2}}\left[(y_0^2-x_0^2)U_2(x,y)+2x_0y_0V_2(x,y)\right].
\end{equation}
The agreement between both calculations is not as good as for the other defect, with a one pixel mismatch, but we keep  this analytical description of the splitting with a value of $h_p^{'}=0.1$ for any tested $r_d/r_0$ ratio.

\subsection{Angular deformation}\label{s_ang}
An angular deformation is more complicated to account for as the frame used to describe the geometry is modified, which has an impact on the analytical expression of other deformations. An angular deformation can always be seen as a combination of the rotation of one quadrupole set compared to the other one and a shearing effect within one or both quadrupole sets.

We start with the shearing effect alone as it does not modify the global orientation of one quadrupole set relatively to the other one (see Fig.~\ref{fig:motif} for a representation). One angle per set ($\beta_S$, $\beta_T$) is required to characterise this defect pattern. In the case of a pure shearing, $\beta_S$ is the angle of the axis passing by the centres of the two electrodes of the $S$-set closest to the $x$-axis of the frame to the $x$-direction, the other trap axis being constrained to have a $-\beta_S$ angle to the $y$-direction of the frame (the angle sign-convention being the trigonometric one). The three aligned local minima induced by such a deformation is the signature of an extra quadrupole term, and the minima configuration can be reproduced by the analytical expression  
\begin{equation}
W_{h}=h_{h}\left[\beta_T U_2(x,y)+\beta_S V_2(x,y)\right]
\end{equation}
with the adjusted parameter $h_h=1.404$  fitting our trap dimensions. The comparison between analytical and numerical results show an agreement  for a shearing angle up to its maximal size $\pi/15$ up to  the smaller pixel size  of $2.6~\mu$m. The coding of this perturbation is then $a_1=h_h\beta_T$, $a_2=h_h\beta_S$, $a_3=a_4=0$

As for a rotation between the two quadrupole sets, defined by the smallest angle $\gamma$ between two neighbour electrodes within the octupole trap, it does not induce any extra minimum in the pseudo-potential. Nevertheless, we see in the next section that it has an impact when combined with other geometric defects as it rotates the alignment of the minima. In that case, the relevant parameter is the angle $\delta$ such that $|\delta|=(\pi/4-\gamma)/2$, oriented positively for a counter-clockwise rotation of the $S$-set. We now consider more realistic situations that combine different basis defects. 

\subsection{A general form for any configuration}
\begin{figure}[htbp]
\begin{center}
\includegraphics[height=6cm]{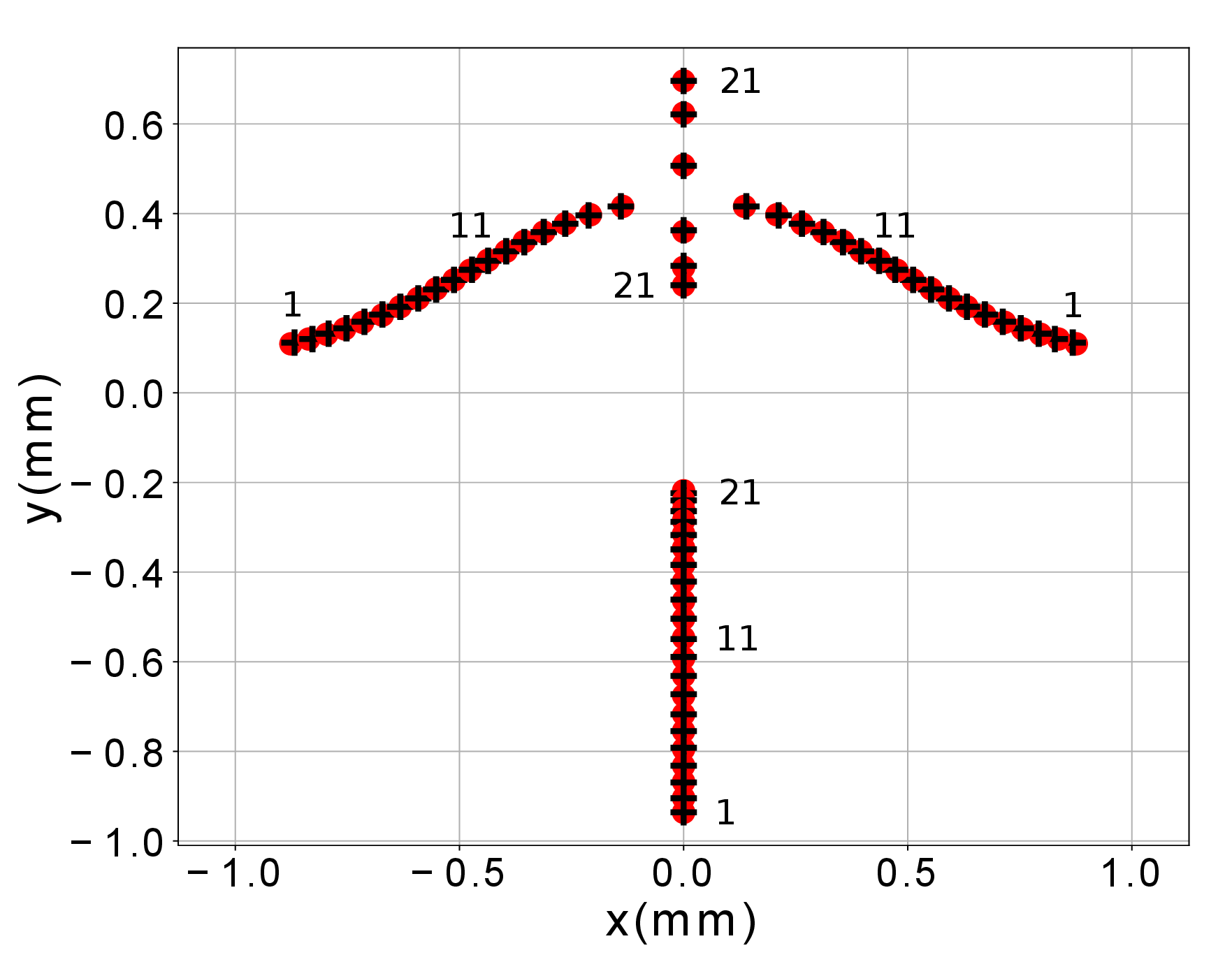}
\caption{Positions in the radial plane of the three minima of the pseudo-potential, computed by CPO (black cross) and by the analytical equation (red dots). They are superposed on the same pixel (size 4~$\mu$m). The trap defects are described by a constant sliding effect within the $S$-quadrupole set, defined by $y_l^{S}=-0.004$, superposed to a compression within the same set, defined by $L_S$ scanned from 0.108 (minima label 1) to -0.111 (minima label 21) with step of size -0.011 }\label{fig_slid_comp}
\end{center}
\end{figure}

In the previous sections, an analytical expression is proposed for five identified defect classes. We propose now to demonstrate that any geometric configuration can be represented as a linear combination of the five identified defects and that their expression can be summed up to analytically describe this configuration. We first focus on situations where the defects do not involve any angular deformation and where the three defects {\it Compression, Sliding} and {\it Splitting} are combined. In that case, there is no cross-effect between the three basic defects and the resulting impact on the potential is a simple linear sum of the contribution of each individual defects. The comparison between the numerical simulations by CPO and the analytical expressions resulting from a simple sum of these three perturbation terms show an agreement  for defects resulting in a mis-position as large as 4\% of the trap radius $(r_0+r_d)$. As an example, we propose to focus on the combination of the dipole-kind sliding within the $S$-quadrupole set, defined by $y_l^{S}=-0.004$ superposed with a quadrupole-kind compression within the same set. The compression parameter $L_S$ is scanned from 0.108 to -0.111 with step of size -0.011. The positions of the three potential minima calculated both by CPO and by the analytical description are shown on Fig.~\ref{fig_slid_comp}. When the compression parameter is scanned, the organisation of the minima starts from a triangle (labelled 1 minima) through a nearly balanced triangle (labelled 11 minima) to a line oriented along the compression and sliding axis (labelled 21 minima). The agreement between the analytical description and the numerical calculations shows the relevance of our method for mis-positioning up to 4\% of the trap radius. 

If we now consider an angular deformation, the total perturbative term is not a simple sum of the basic ones and the analytical expression of each perturbative term must be modified to represent accurately the trap configuration. For example, to account for the rotation $\delta$ of one quadrupole set compared to the other one (like defined in \ref{s_ang}), the quadrupole terms, $U_2$ and $V_2$ have to be corrected in the expression of a compression $W_c$, and shearing $W_{h}$ perturbation. We define the matrix 
\begin{equation}
\overline{R(\theta)}=\left( \begin{array}{cc}
\cos(\theta) & \sin(\theta) \\
\sin(\theta) & \cos(\theta) \end{array} \right)
\end{equation}
which is not a conventional rotation matrix because it acts on quadrupole terms. In case of a {\it compression+rotation} configuration, $W_c$ becomes $W_c^R$ which can be analytically expressed by changing $(U_2(x,y), V_2(x,y))$ into $(U_2^c(x,y),V_2^c(x,y))$ such that $(U_2^c ;V_2^c)=\overline{R(-2\delta)}(U_2 ;V_2)$. In case of a {\it shearing+rotation} configuration, $W_{h}$ becomes $W_{h}^R$ with $(U_2^{h} ;V_2^{h})=\overline{R(\delta)}(U_2 ;V_2)$. For a {\it sliding} defect, the rotation effect is already included in the definition and the {\it splitting} description does not need an axis rotation to remain accurate but we observe a rescaling of the perturbative contributions of the three non-angular defects. Altogether, the perturbation terms adding up to the octupole instantaneous potential write
\begin{eqnarray}
W^R(x,y)&=&\left(1+\frac{|\delta|}{\pi}\right)W_{h}^R(x,y)   \nonumber \\
&+&\left(1+\frac{2|\delta|}{\pi}\right)\left[W_c^R(x,y)+W_{l}(x,y)+W_{p}(x,y)\right]
\end{eqnarray}
with $W_{h}^R(x,y)=h_{h}\left[\beta_T U_2^h(x,y)+\beta_S V_2^h(x,y)\right]$, $W_c^R(x,y)=h_c\left[L_SU_2^c(x,y)+L_TV_2^c(x,y)\right]$ and $W_{l}(x,y)$ and $W_{p}(x,y)$ remains as in Eq.~\ref{eq_Sd} and~\ref{eq_Spb}.

The last step is to combine a shearing effect to the other deformations : in the case of a {\it shearing+compression} defect, the three potential minima still align around the trap center but the modification in the inner radius has an impact on the scaling of the perturbation terms. More precisely, it is the difference $\Delta r$ between the inner radius in the $T$-quadrupole set and the $S$-quadrupole set that counts and $W_{h}$ must be coded like $a_1=h_h\beta_T(1-3\Delta r/r_0)$, $a_2=h_h\beta_S(1+3\Delta r/r_0)$, $a_3=a_4=0$.

In the case of a {\it shearing+sliding} defect, the minima organise on a triangle and the analytical expression depends on the involved quadrupole sets. Indeed, if these two defects are sported by the same set, the projection procedure described in \ref{sliding} accounts for the shearing effect and there is no need for an extra calculation step. In the other case, the definition of the sliding parameter depends on the shearing of the other set. As an example, if the $T$-set is sheared by $\beta_T$ and the sets are deformed by a sliding defined by ($x_{l}^S,y_{l}^S$) and ($x_{l}^T,y_{l}^T$), the perturbation term $W_{l}$ must be coded as $a_1=a_2=0$, $a_3=h_l((1+4\beta_T/\pi))x_{l}^S+x_{l}^T)$ and $a_4=h_l((1-4\beta_T/\pi))y_{l}^S+y_{l}^T)$. 

As for a combination of {\it shearing+splitting} defect, the minima also organise as any triangle and the calculated potential matches the numerical calculation if the perturbation $W_{p}$ is now coded by $a_1=a_2=0$, $a_3=h_p(1+\sin(2\beta_T)\cos(2\beta_S)x_0-\sin(2\beta_S)y_0)$ and $a_4=h_p(1-\sin(2\beta_T)\cos(2\beta_S)y_0-\sin(2\beta_S)x_0)$. The equation summing all the corrected contributions is collected in \ref{a_global}.

\begin{figure}[htbp]
\begin{center}
\includegraphics[height=6cm]{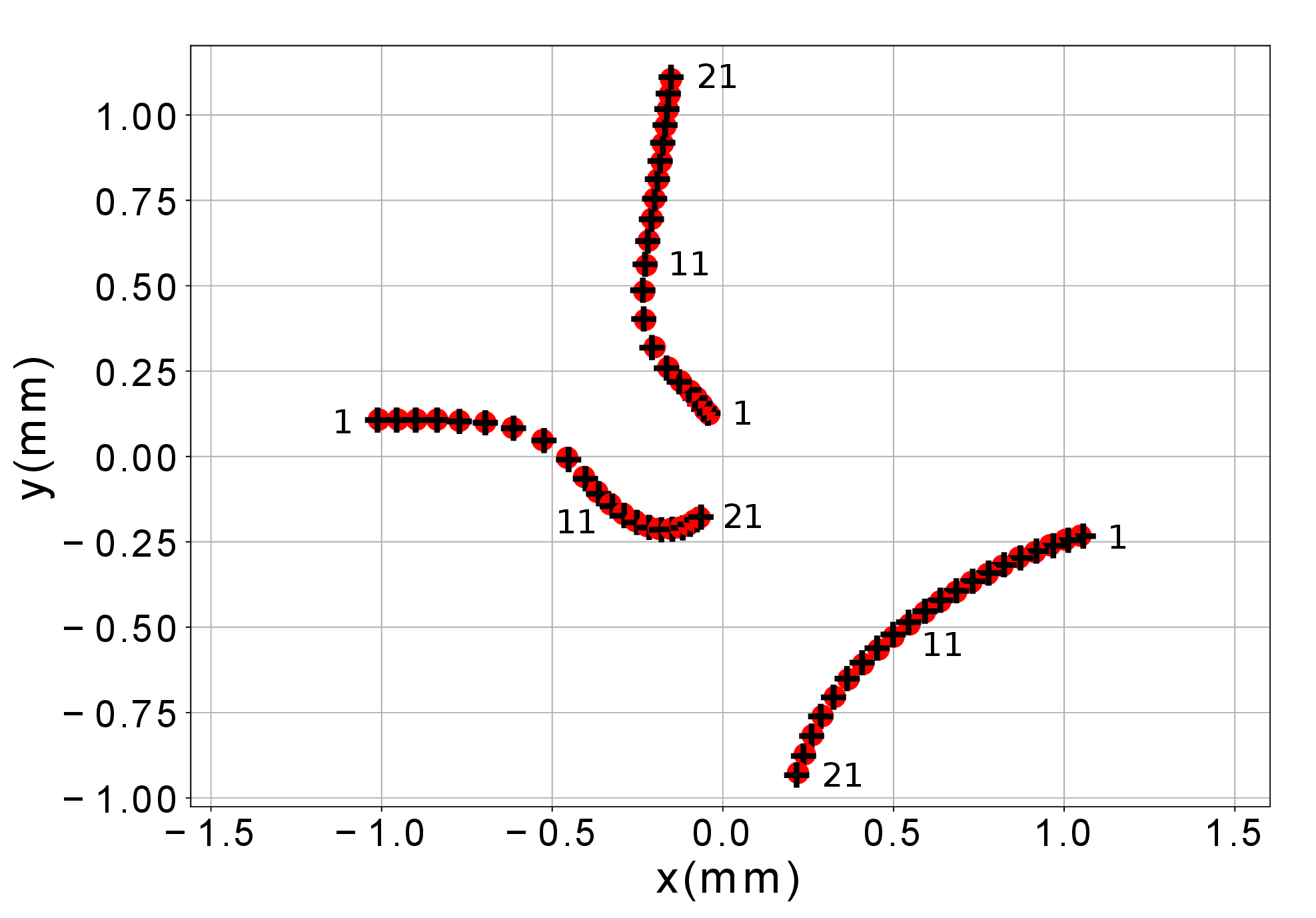}
\caption{Positions in the radial plane of the three minima of the pseudo-potential, computed by CPO (black crosses) and by the analytical equation (red dots). The electrode position defects are described by a fixed splitting effect defined by $x_0=0.007$, superposed to a compression within the $T$-quadrupole set, defined by $L_T=0.55$ and a shearing within the same set scanned from $\beta_T=0.088$  (minima label 1) to $\beta_T=-0.088$  (minima label 21) with step of size -0.009  }\label{fig_split_comp_shear}
\end{center}
\end{figure}

To illustrate a general case, we propose on Fig.~\ref{fig_split_comp_shear} to observe this matching for electrode positions that can be described by a compression ($L_T=0.55$)+splitting ($x_0=0.007$)+shearing with a shearing parameter $\beta_T$ scanned from 0.088 to -0.088. Once again, the analytical description and  the numerical calculations agree within the 4~$\mu$m pixel size.

\subsection{Completeness of the defect basis}
In the previous sections, agreement between numerical and analytical results were demonstrated for designed geometric configurations. To confirm the relevance of the description, we propose to check that any random electrode configuration can be mapped onto the 5 basic defects. To that purpose, we  decomposed  200 electrode configurations onto the five defects basis. These configurations were built such that the positions of the centers of the eight electrodes are randomly chosen at a distance from their ideal positions ranging 0.5\% to 4\% of the distance to the trap centre $r_0+r_d$. The comparisons between numerical and analytical results are ordered by the average distance $\overline{d}$ between the local potential minima calculated with CPO and with the analytical description. The resolution is limited by the 4~$\mu$m pixel size relatively to the distance $r_0+r_d=5.5$~mm. To compare very different potential minima configuration, we rescale the average distance $\overline{d}$ relatively to the mean distance $d_b$ between the three minima and their mass center, and write it as $\overline{d}_s$. For a position mismatch of each electrode equal to 1\% of $r_0+r_d$, 182 (respectively 198) cases out of 200 have a mean relative distance $\overline{d}_s$ lower than 2\% (respectively 4\%). These ratios do not change much when the mis-positioning is 2\% of $r_0+r_d$. When it reaches 4\%, these numbers decrease to 132 cases out of 200 that have a mean distance $\overline{d}_s$ lower than 2\% and 173 cases with a mean distance lower than 4\%. This very good statistics demonstrate that the description in terms of defect basis is relevant to reproduce the potential minima configuration for defect size of few \% of the trap size. This limit is reachable to machined macroscopic trap of millimetre size.  

\section{Control of the RF voltages}\label{s_cor}

In this last section, we show how the analytical description that we propose above allows the calculation of a set of relative RF-amplitude that must be applied to compensate for the geometric defects. Then, starting from a compensated potential, on can create designed line or triangle minima configurations. The description of identified defects as dipole- or quadrupole-kind shows that several geometrical configurations can be responsible for the same potential minima configuration. As an example, a compression described by $L_s=0.0554$, which corresponds to a 2\% positioning error relative to $r_0+r_d$, creates the same pattern as a shearing described by $\beta_T=0.0319$, which corresponds to an angular mis-positioning of 0.032 rad (with position resolution of $2\times 10^{-3}r_0$). Our interest being in the shape of the potential itself, we developed a strategy that bypass the identification of the geometric defects and the positions of the electrodes as there is no one-to-one connection with a minimum configuration. 

\subsection{Producing controlled patterns in a perfect octupole}

The  perturbation induced by the structure being a sum of quadrupole and dipole terms, we  deduce  a set of applied voltages on the electrodes that generate quadrupole and dipole perturbations in the potential. The compensation strategy is based on the identification of the $a_{1,2,3,4}$ coefficients involved in the analytical equation to apply the appropriate counter-perturbation $-a_{1,2,3,4}$ through custom applied voltages on the electrodes. The method we propose is to map a $a_{1,2,3,4}$ set to a set of voltage bias $\delta V_k$ added to the average RF-voltage on each electrode, with $k$ its index. Applying $-\delta V_k$ to the trap electrodes should, in principle, compensate for the geometric defects. This is only partly verified because the $-\delta V_k$ are applied on a trap showing a perturbed geometry but we choose this strategy as a starting point. 

There are different $\delta V_k$ patterns that can be applied to produce the same perturbation in the potential, and one is given here that presents the advantage of having the most accurate representation of dipole terms and sharing the compensation between the largest number of electrodes :
\begin{eqnarray}
\delta V_0 = V_{rf}\left(+\frac{a_1}{q_c} - \frac{a_4}{d_c}\right); \ \delta V_1 &=& V_{rf}\left(- \frac{a_2}{q_c} - \frac{1}{\sqrt{2}}\frac{a_3+a_4}{d_c}\right); \nonumber \\
\delta V_2 = V_{rf}\left(- \frac{a_1}{q_c} - \frac{a_3}{d_c}\right); \ \delta V_3 &=& V_{rf}\left(+\frac{a_2}{q_c} - \frac{1}{\sqrt{2}}\frac{a_3-a_4}{d_c}\right); \label{eqn:coeff_to_detuning} \\
\delta V_4 = V_{rf}\left(+\frac{a_1}{q_c} + \frac{a_4}{d_c}\right); \ \delta V_5 &=& V_{rf}\left(- \frac{a_2}{q_c} +\frac{1}{\sqrt{2}}\frac{a_3+a_4}{d_c}\right); \nonumber \\
\delta V_6 = V_{rf}\left(- \frac{a_1}{q_c} + \frac{a_3}{d_c}\right); \ \delta V_7 &=& V_{rf}\left(+\frac{a_2}{q_c} +\frac{1}{\sqrt{2}}\frac{a_3-a_4}{d_c}\right); \nonumber
\end{eqnarray}
The electrodes are numbered clockwise, starting by the bottom. The linear dependency of the $\delta V_k$ with the $a_{1,2,3,4}$ was established through numerical simulations by generating potentials according to Eq.~ \ref{eq_W} with varying $a_{1,2,3,4}$ coefficients and collecting the positions of the minima as reference. The CPO software was then used to generate the potential map of a perfect octupole trap ($(r_0,~r_d)$=(4, 1.5)~mm) with individually tuneable potentials on the electrodes. The potentials on the electrodes of the simulated octupole trap were applied according to Eq.~\ref{eqn:coeff_to_detuning}, with adjusted values of the $d_c$ and $q_c$ calibration coefficient to ensure the superposition of the positions of the  minima of the pseudo-potential deduced from Eq.~(\ref{eq_U},\ref{eq_W}) and from the CPO simulations. 

\begin{figure}[!h]
\centering
\includegraphics[height=7cm]{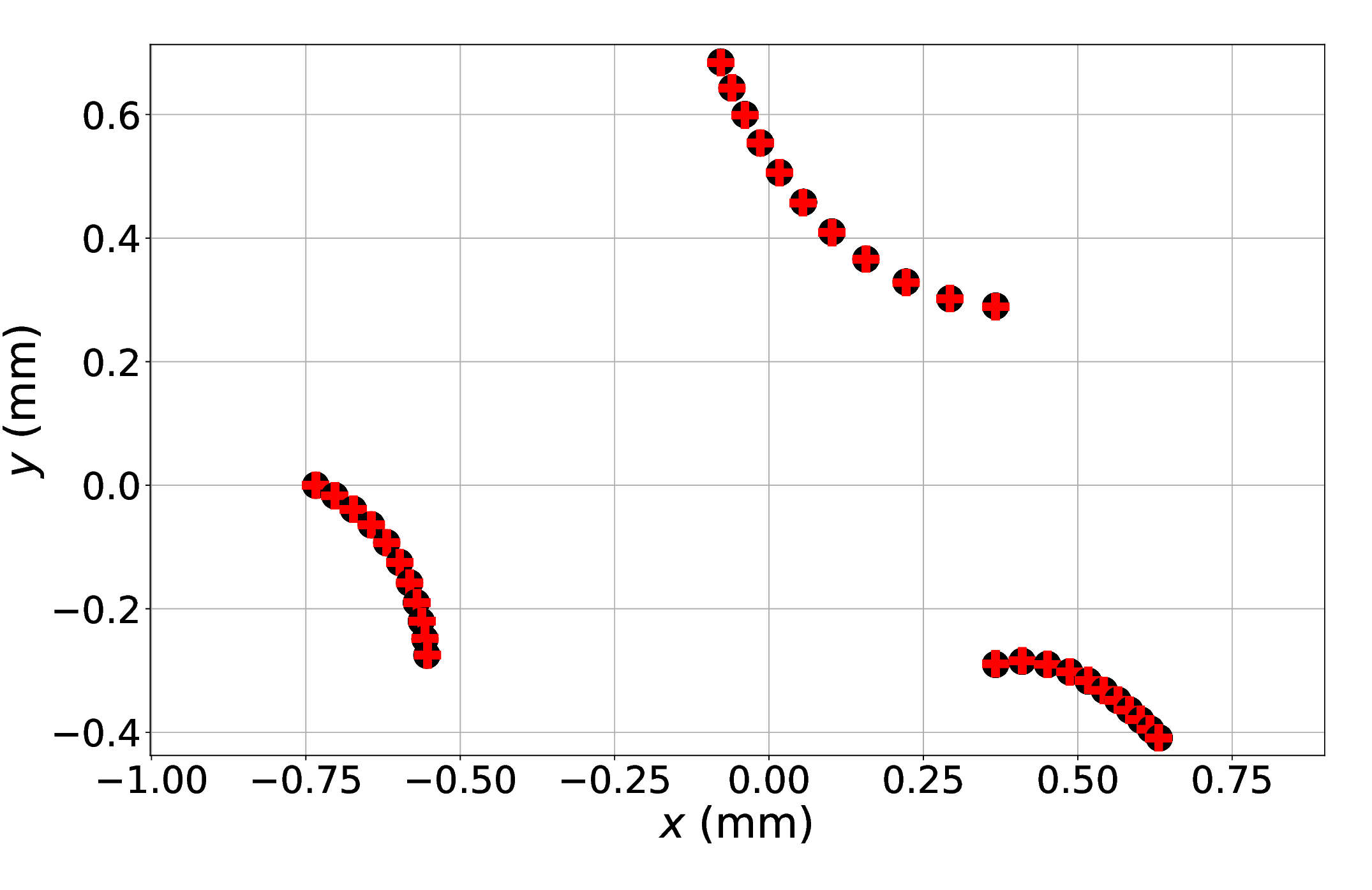}
\caption{Red crosses: positions of the minima in the pseudo-potential generated from Eq.~ \ref{eq_Spb} for $a_{1,2,3,4}$ coefficients following the dependency : $\{0.4(0.1-j),~0.2j,~-0.1(0.1-j),~0.2j\}$ with $j$ spanning from 0 to 0.1 with steps of 0.01. Black dots: corresponding positions of the minima for the simulated perfect octupole trap with the calculated RF voltages $V_k = V_{rf}+\delta V_k$ according to Eq.~ \ref{eqn:coeff_to_detuning}. The size of the grid in both cases is $d_{px}=$1~$\mu$m.}
\label{fig:gen_detu_cal}
\end{figure}

With this protocol we find $d_c=0.912\pm 0.008$ and $q_c=0.796 \pm 0.008$, the uncertainty comes from the range of values that allows for a superposition at the pixel size which is reduced to 1~$\mu$m to improve the precision. The fit was ensured for values of $a_i$ up to $0.1$ which, for the sake of comparison, is a little larger than the amplitude expected for a single involved deformation class reaching of 4\% of the $(r_0+r_d)$ distance if the potential deformation were of mechanical origin. This ought to ensure we are not out of range of the validity domain of this fit when we apply the correction scheme in our trap where the defect amplitude is expected to be about 2\% of the $(r_0+r_d)$ distance at most. Fig.~\ref{fig:gen_detu_cal} shows a fit to the pixel size between minima generated from the surface equation Eq.~\ref{eq_W} for varying $a_{1,2,3,4}$ coefficients and the minima generated in the perfect octupole trap by converting the $a_{1,2,3,4}$ coefficients into adapted RF voltages according to Eq.~\ref{eqn:coeff_to_detuning}. The selected tuning patterns allows the controlled generation of any perturbation in the octupole trap that writes as a sum of quadrupole and dipole terms. 

\subsection{Diagnosing the perturbations from the positions of the minima}
In order to compensate the perturbations of the potential by applying the proper counter-perturbation through a set of adapted voltages, it is necessary to establish a diagnosis of the perturbations from experimentally accessible data. Working directly with the perturbations of the potential instead of the deformation of the structure presents the advantage of reducing the number of unknown parameters from 16 to five ($a_{1,2,3,4}$ and $\bar{r_0}$). In the context of an experiment, we can bypass the knowledge of the positions of the electrodes to find the $a_i$ coefficients by working directly with the relative positions of the three potential minima in the octupole frame, assuming the trapped ions bunch in these minima once laser cooled \cite{pedregosa18}. The set of parameter ($a_{1,2,3,4}$, $\bar{r_0}$) being unique to a given perturbation of the potential, we use the position of the minima to evaluate the parameter set $a_{1,2,3,4}$ by assuming a value for $\bar{r_0}$, usually $\bar{r_0}=r_0$. Starting from guess initial values for the $\{a_i\}$, the pseudo-potential minima positions generated from Eq.~\ref{eq_W} are compared to the observed minima. Like in \cite{pedregosa18}, the figure of merit is the average of the three distances $\bar{d}$ between the calculated and the measured minima and an optimisation routine looks for the set of $a_i$ coefficients which minimises this figure. 

To validate the possibility to deduce the perturbations from the positions of the minima we have generated a number of test cases with known perturbations and applied the search protocol to see if the returned coefficients correspond to the inputs values, with an expected result $\bar{d}=0$. To evaluate the impact of the resolution on the positioning of the minima, determined in practice by the collection optics, the calculated pseudo-potential is discretised with three different pixel sizes ($d_{px}=$~2~$\mu$m, 4~$\mu$m and 8~$\mu$m). The positions of the minima were collected from calculated potentials with randomised $a_i$ coefficient with realistic order of magnitude, which corresponds to $\pm 0.1$ for a 4\% amplitude on the defects and $\pm 0.05$ for a 2\% case. In our tests, coefficients can have any value in the interval $[-0.1,0.1]$. 50 randomised cases were tested for each pixel sizes. In regard to the sampling of the surface, a success condition for the search protocol can be set to $\bar{d}< d_{px}$ or $\bar{d}<2\times d_{px}$. The success rate of the search protocol is summarised in Table.~\ref{tab:success_rate_code}. 
\begin{table}[h!]
\begin{center}
\begin{tabular}{c|c|c}
&Success rate if :&Success rate if : \\
$d_{px}$ ($\mu$m) & $\bar{d}< d_{px}$ (\%) & $\bar{d}<2\times d_{px}$ (\%) \\
$2$ & 78 & 96 \\
$4$ & 84 & 92  \\
$8$ & 66 & 90  \\
\end{tabular}
\caption{Success rate of the code predicting the $\{a_i\}$ coefficients pondering the perturbations from the positions of the minima in the resulting pseudo-potential.}
\label{tab:success_rate_code}
\end{center}
\end{table} 
To estimate the closeness of the $\{a_i\}$ coefficients returned by the code to the input coefficients, we use the standard deviation $\sigma_{1,2,3,4}$ of their difference  over the cases fulfilling the success condition $\bar{d}< d_{px}$. The results are summarised in Table.~\ref{tab:prec}, and show that a smaller pixel size leads to a better  accuracy on the estimated coefficients. We validate our search protocol as adequate to the purpose of identifying the perturbations in the potential from the position of the minima in the frame of our compensation protocol. 
\begin{table}[h!]
\begin{center}
\begin{tabular}{c|c|c|c|c}
$d_{px}$ ($\mu$m) & $\sigma_1 $ & $\sigma_2$ & $\sigma_3$ & $\sigma_4$ \\
$2$ & 0.000293 & 0.000370 & 0.000223 & 0.000249 \\
$4$ & 0.000820 & 0.000878 & 0.000470 & 0.000420 \\
$8$ & 0.001787 & 0.001785 & 0.000973 & 0.000757 \\
\end{tabular}
\caption{Standard deviation of the difference between the input coefficients and the coefficients estimated from the search protocol, for the ‘successful cases’ fulfilling the condition $\bar{d}< d_{px}$.}
\label{tab:prec}
\end{center}
\end{table} 

\subsection{Compensation of geometric defects}
\begin{figure}[!h] 
\centering
\includegraphics[width=0.7\textwidth]{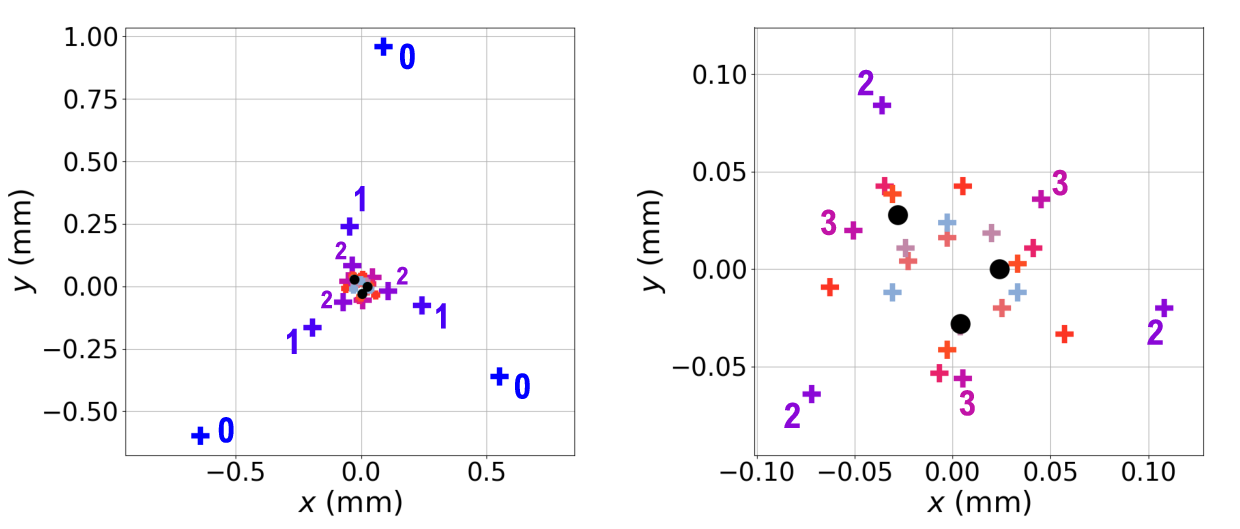}
\caption{Positions of the pseudo-potential minima in the radial plan for ten step of the iterative correction protocol, one color per step, on a structure with 2\% random error on the position of the electrodes. The right panel is a zoom on the central region of the left panel. The black dots correspond to the position of the minima at the end of the tenth iteration of the protocol.}
\label{fig:min_step_conv}
\end{figure}

\begin{figure}[!h] 
\centering
\includegraphics[width=0.5\textwidth]{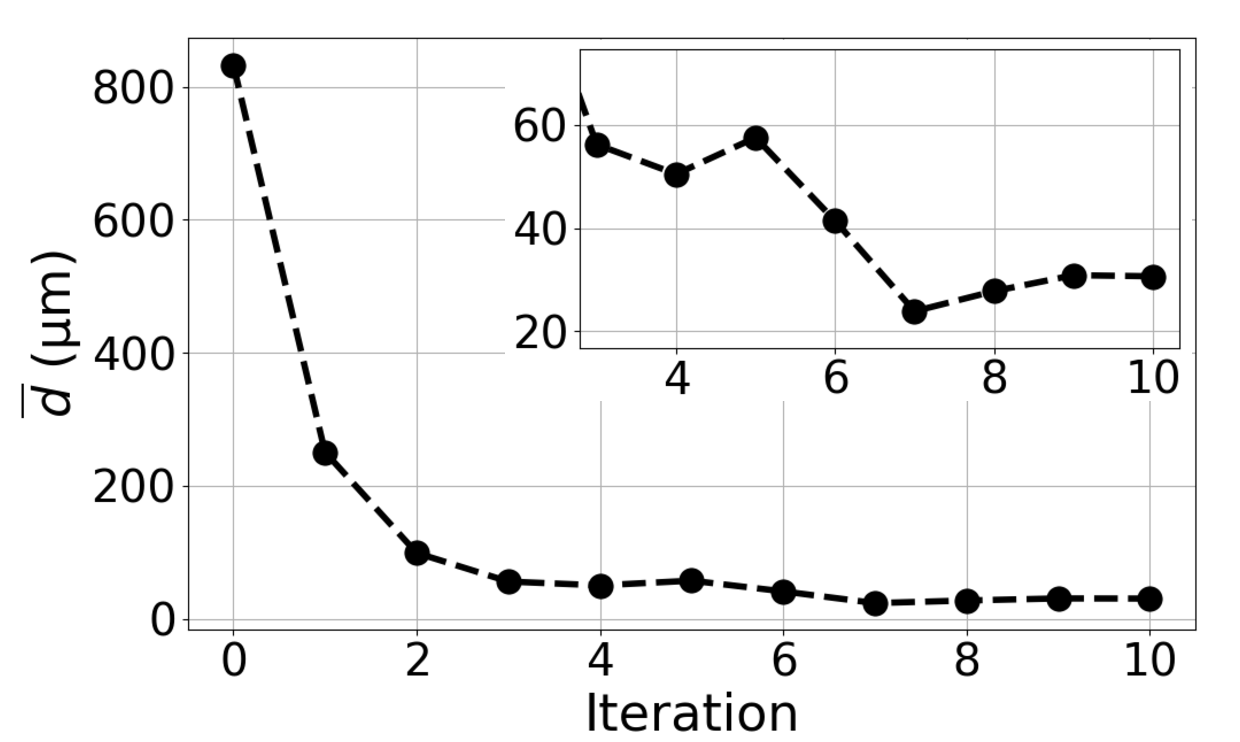}
\caption{Average distance $\bar{d}$ of the minima to their barycentre calculated from the position plotted in Fig.~\ref{fig:min_step_conv}, for each step of the iterative correction protocol.}
\label{fig:dist_step_conv}
\end{figure}

To test the compensation in an octupole trap affected with structural deformations, we impose that the generated geometries are free from a general rotation. From the positions of the minima we can diagnose the weights $a_{1,2,3,4}$ characterising the structure induced perturbation of the potential and with the set of equations~(\ref{eqn:coeff_to_detuning}) we can apply the appropriate counter-perturbation characterised  by $-a_{1,2,3,4}$ by tuning the voltages on the electrodes to compensate for the structure deformations. Simulations have shown that  compensation of the potential is not achieved after one step of this protocol, which is due to the deformation of the structure modifying slightly the value of the applied counter-perturbation in the potential. Fig.~\ref{fig:min_step_conv} shows a practical example in an octupole trap with a random displacements of the electrodes with a norm of 2\% of the $(r_0+r_d)$ distance : the positions labelled 0 correspond to the positions of the minima before compensation is applied, and their average distance to their barycentre is $\bar{d}_b=$~830~$\mu$m. After a first correction step, the minima, labelled 1, are brought closer to the centre but not merged since they are still distant of $\bar{d}_b=$~250~$\mu$m (see Fig.~\ref{fig:dist_step_conv}). 

By repeating the correction protocol several times, each iteration ought to allow the potential to be brought closer to the ideal shape by getting finer information on the tuning voltages. More precisely, at each step $j$ the correction protocol goes as follow: the positions of the minima are gathered from the fluorescence of the ions in the trap. The associated coefficient set $\{a_i\}$ characterising the present potential deformations is determined via the search protocol. A correction set $\{a_i\}_j$  is calculated and used to update the correction voltages  by building up with the previous values and the total correction voltage on each electrode is cumulative with the previous steps. To illustrate the step by step functioning of the correction protocol, ten iterations of the correction protocol are applied on the test trap. Fig.~\ref{fig:min_step_conv} shows the positions of the minima in the radial plane for each correction step, and the average distance of the minima to their barycentre is plotted on Fig.~\ref{fig:dist_step_conv}. The first three correction steps reduce the average distance between the minima of 93.5\% and the next steps further reduce it of another 3\%. Given the pixel size of 4~$\mu$m, the average distance between the minima reaches its limit  at the seventh iteration, with a  value of 24~$\mu$m. Iterating the protocol further rearranges the positions of the minima but does not bring them closer to merging.

Two parameters govern the achievable quality of the compensation: they are the pixel size, and the allowed resolution on the coefficients $a_i$. In an experiment they are limited by the resolution of the optics  and the ion motion for the positioning of the minima, and the available resolution on the tuning voltages  $\delta V_k$ (see Eq.~\ref{eqn:coeff_to_detuning}). The optimum scenario for the compensation was  identified  for pixel sizes of 4~$\mu$m, and a resolution of $10^{-6}$ on the $a_i$. In both cases there is a notable deterioration of the results for bigger pixel sizes and lower resolutions, but no notable gain is achieved with more stringent requirements. With this pixel size and resolution, five iterations  were sufficient, on average, to see no further gain in the average distance $\bar{d}_b$ over 10 test cases with a 2\% perturbation range. The average distance between the minima keeps around 40~$\mu$m on average, which means that even in the best-case parametrisation the spatial merging of the minima is not achieved. Nevertheless, regarding their potential depth, they are reduced to a few microkelvins compared to their initial depths of the order of a few hundreds of kelvins. The microkelvin scale is negligible  in regard of the temperature of laser cooled ions which is of the order of 10~mK and we can consider that the three local minima in the potential are compensated after five iterations.

So far, the  contribution of the DC-deconfining term induced by the DC-trapping along the trap axis \cite{drewsen00} has been neglected in the protocol even if its impact on the trapping potential shape can not be avoided. By turning a flat bottom well into  ring shaped geometry, it smears the three potential minima into two or even one, setting a limit to the compensation protocol that stops before the best estimated performance are reached \cite{pedregosa18}. The efficiency of the proposed method is then dependent on the maximum RF amplitude and minimal DC-voltage that the set-up can reach to observe ions in the most favorable configuration for a relevant compensation. Once the tuning parameters identified, the operating voltages can be modified as the compensation voltages can be adapted accordingly. With a DC-voltage added to the RF-pseudo-potential, it was found in \cite{pedregosa18} that a relative resolution of $10^{-4}$ on $\delta V_k$ was sufficient to reach the best reachable performances, with residual potential variation along the ring of the order of few millikelvins. 

\section{Conclusion}

The description of a realistic linear octupole trap proposed in this paper is based on the adjunction of lower order contributions in the RF potential created by the electrodes. The protocol for diagnosing of the impact of structural defects on the potential and their compensation relies on the positioning of the minima in the RF-induced pseudo-potential through the imaging of the trapped cold ions by their laser induced fluorescence. This protocol was motivated by the practical realisation of the compensation of the potential asymmetries introduced by structural defects, to form a trapping potential with a shape as symmetric as possible. In practice, it is limited by the impact of the DC-deconfining contribution in the radial plane that modifies the ion organisation when they are few. In the large number limit, they organise independently of this contribution \cite{champenois09} and another strategy needs to be found to go further in the defect compensation. Nevertheless,  the proposed description is a tool that allows for the control design of three ion chains \cite{marciante11}, a relevant configuration to simulate and study 3D frustrated spin system like  demonstrated in  \cite{friedenauer08} for 2D system where the transition from paramagnetic to ferromagnetic order was observed for 3 ions.

\section*{acknowledgement}
 MM acknowledges DGA for funding and stimulating scientific environnement. This experiment has been financially supported by CNES (contract 151084).

\appendix
\section{Geometrical representation of the defect basis set}\label{a_motif}
\begin{figure}[!h] 
\centering
\includegraphics[width=0.7\textwidth]{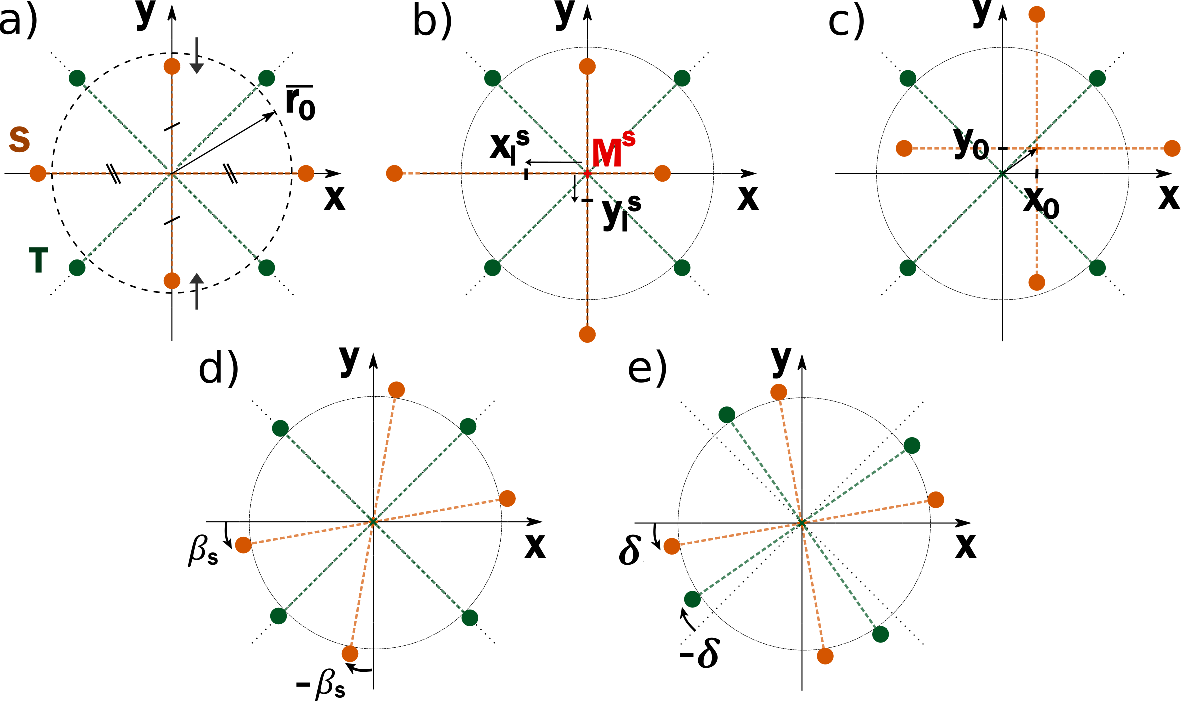}
\caption{Representation of the parameters defining a defect for a) : Compression, applied to the $S$ electrode set,   b) : Sliding, in the case of a sliding in the $S$ electrode set,   c): Splitting,  d): Shearing, applied to the $S$ electrode set,   and e) : Rotation. The dotted line circle is at a distance $r_0$ from the trap center.}
\label{fig:motif}
\end{figure}

Figure~\ref{fig:motif} gives a representation of each defect of the basis set in some specific situations, detailed in its caption.

\section{Complete equation for a configuration involving the five identified defects} \label{a_global}
For a random configuration of electrodes, the $\{a_i\}$ coefficients result from the sum of the contribution from the five basis defects, modified by an eventual angular deformation, as described in section~\ref{s_pert_list}. The following equations give the full analytical expression for these coefficients.
\begin{eqnarray}
a_1 &=& h_c\left(1+\frac{2|\delta|}{\pi}\right)\left(L_S\cos(2 \delta)-L_T\sin(2\delta)\right) \nonumber \\
&+& h_h\left(1+\frac{|\delta|}{\pi}\right)\left(\beta_T\left(1-3\frac{\Delta r}{r_0}\right)\cos(\delta)+\beta_S\left(1+3\frac{\Delta r}{r_0}\right)\sin(\delta)\right)\\
&-& 0.1\frac{x_0^2-y_0^2}{\sqrt{x_0^2+y_0^2}} \nonumber \label{eqn:full_coeff_set_per1} \\ 
\nonumber \\
a_2&=& h_c \left(1+\frac{2|\delta|}{\pi}\right)\left(L_T\cos(2 \delta)-L_S\sin(2\delta)\right) \nonumber \\
&+& h_h\left(1+\frac{|\delta|}{\pi}\right)\left(\beta_T\left(1-3\frac{\Delta r}{r_0}\right)\sin(\delta)+\beta_S\left(1+3\frac{\Delta r}{r_0}\right)\cos(\delta)\right) \\
&+& 0.1\frac{2x_0y_0}{\sqrt{x_0^2+y_0^2}} \nonumber \\
\nonumber \\
a_3&=& [h_l\left(\left(1+4\frac{\beta_T}{\pi} \right)x^S_{l}+x^T_{l}-4\frac{\beta_T}{\pi}y^T_{l}\right)\nonumber \\
&+& h_p (1+\sin(2\beta_T)\cos(2\beta_S))x_0-\sin(2\beta_S)y_0]\left(1+\frac{2|\delta|}{\pi}\right) \\
\nonumber \\
a_4&=& [h_l\left(\left(1-4\frac{\beta_T}{\pi} \right)y^S_{l}+y^T_{l}-4\frac{\beta_T}{\pi}x^T_{l}\right) \nonumber \\
&+& h_p (1-\sin(2\beta_T)\cos(2\beta_S))y_0-\sin(2\beta_S)x_0]\left(1+\frac{2|\delta|}{\pi}\right)
\label{eqn:full_coeff_set_per4}
\end{eqnarray} 

\section{Scaling the perturbation terms to any radius sizes}\label{s_rayon}
\begin{figure}[htbp]
\begin{center}
\includegraphics[height=4cm]{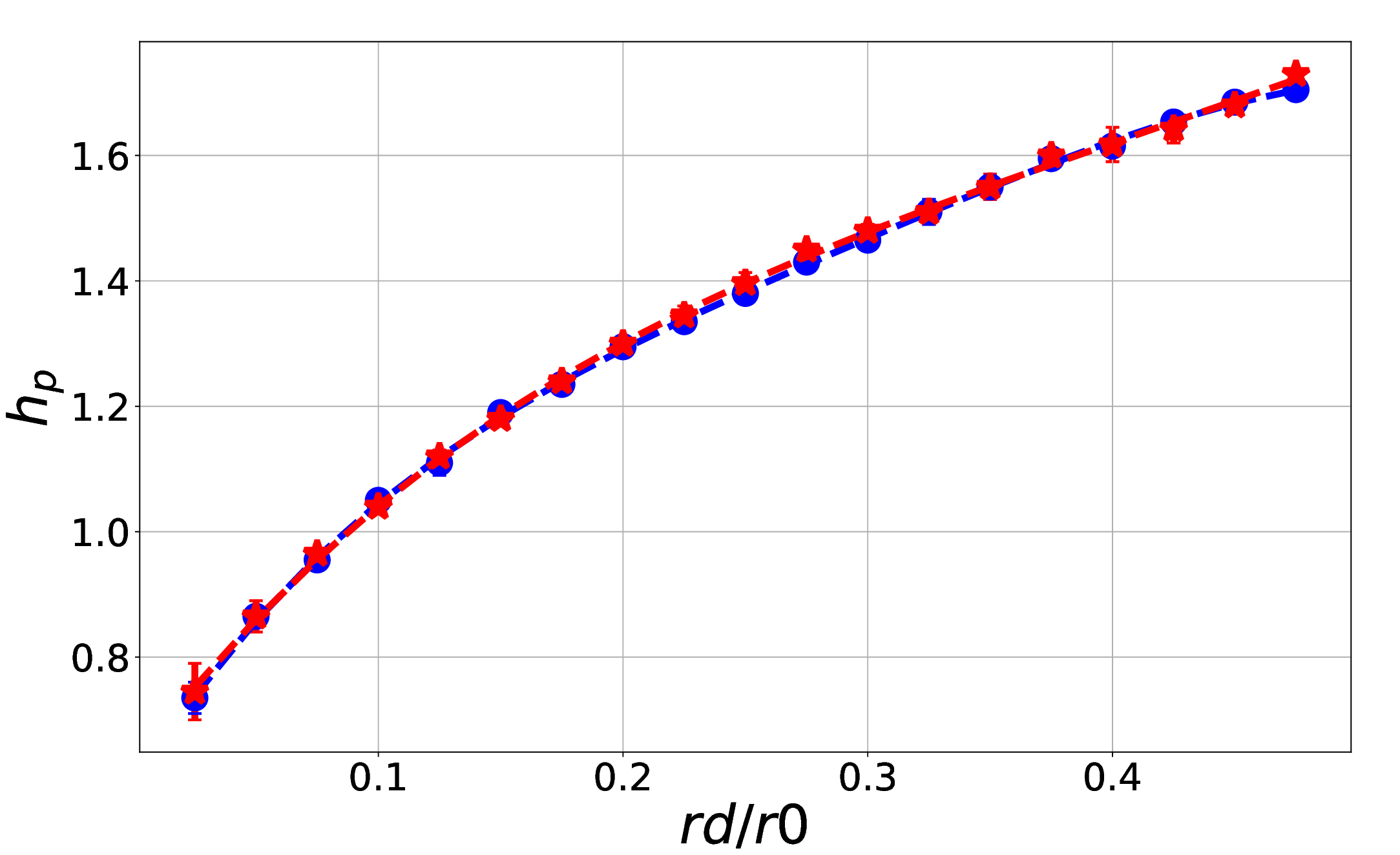}
\caption{Coefficient $h_p$ scaling the splitting perturbative term vs the trap radius ratio $r_d/r_0$. The blue and red points correspond to a splitting defect of  respectively $y_0=0.055$ and $y_0=0.0275$  which corresponds in the case of an octupole with $(r_d, r_0)=$(1.5~mm, 4~mm) to a distance between the S and T-sets of  $110~\mu$m and $220~\mu$m (2\% and 4\% of the $r_d+r_0$ total length). The $r_d/r_0$ error bar is due to the $4~\mu$m pixel size of the simulated CPO surface. The $h_p$ error bar covers the different values  that allow us to reproduce the position of the minima within one pixel. The continuous lines are the best fitted polynomial curves. }
\label{fig_h}
\end{center}
\end{figure}

The comparisons between analytical expressions and numerical calculations shown in the previous section where done for $r_d=1.5$~mm and $r_0=4$~mm and we stated that the radius scale effects were all taken into account through the scaling constants $h_{0,c,l,p,h}$.  The correlations between several numerical calculations run with 1~mm$<r_0<$5~mm for $r_d=0.5$~mm and 0.1~mm$<r_d<$1.9~mm for $r_0=4$~mm show that these constants only depend on the ratio $r_d/r_0$. For  configurations with only one basic defect introduced in the electrode geometry, the dependence of constants $h_{0,c,l,p,h}$ with this ratio is computed by scanning $r_d$ from $0.02 r_0$ to $0.55 r_0$ for $r_0=4$~mm. An exemple of such a numerical study is shown on Fig.~\ref{fig_h} for two splitting defects. The agreement between the $h_p$ coefficient for two amplitudes of the defect confirms that they scale for $r_d/r_0$ only.

Within the $4~\mu$m pixel size, the dependence of each scaling constant with $r_d/r_0$ can be fitted by a polynomial equation like 

\begin{eqnarray*}
h_0 &=& -0.4557 -7.028(r_d/r_0) +53.7(r_d/r_0)^2 -254(r_d/r_0)^3 \dots  \\
    &    &  +687(r_d/r_0)^4-973 (r_d/r_0)^5+ 556 (r_d/r_0)^6  \\
h_c&=& +0.565+1.138(r_d/r_0)-2.073 (r_d/r_0)^2+3.021(r_d/r_0)^3-1.995(r_d/r_0)^4 \\
h_l&=& +1.141+4.869 (r_d/r_0)-5.880 (r_d/r_0)^2+10.696(r_d/r_0)^3-6.975(r_d/r_0)^4\\
h_p&=&+0.614+5.639(r_d/r_0)-16.260(r_d/r_0)^2+29.980(r_d/r_0)^3-22.084(r_d/r_0)^4 \\
h_{h}&=& +4.208+4.989(r_d/r_0)-3.753(r_d/r_0)^2 -0.0025(r_d/r_0)^3+2.1279(r_d/r_0)^4 \\
\end{eqnarray*}


\end{document}